\documentclass[namedreferences]{SolarPhysics}
\usepackage[optionalrh]{spr-sola-addons} 
\usepackage{graphicx}        
\usepackage{url}
\newcommand{\etal}{{\it et al.}}
\begin{document}                                                                  
\begin{article}
\begin{opening}
\title{Zonal  Velocity  Bands and the Solar Activity Cycle}
\author{K.~R.~\surname{Sivaraman}$^1$\sep H.~M.~\surname{Antia}$^2$\sep
S.~M.~\surname{Chitre}$^3$\sep V.~V.~\surname{Makarova}$^4$}
\institute{$^1$ Indian Institute of Astrophysics, Bangalore 560034, India\\
$^2$ Tata Institute of Fundamental Research, Homi Bhabha Road, Mumbai 400005,
India,
email: \url{antia@tifr.res.in}\\
$^3$ Centre for Basic Sciences, University of Mumbai, Mumbai 400098, India\\
$^4$ Kislovodsk Solar Station of the Pulkovo Observatory, Kislovodsk 357700,
Russia}
\runningauthor{K.R. Sivaraman {\it \etal}}
\runningtitle{Zonal Velocity Bands and the Solar Activity Cycle}

\begin{abstract}
We compare the zonal flow pattern in subsurface layers of the Sun with
the distribution of surface magnetic features like sunspots and polar
faculae. We demonstrate that in the activity belt, the butterfly
pattern of sunspots coincides with the fast stream of zonal flows, although
part of the sunspot distribution does spill over to the slow stream. At high
latitudes, the polar faculae and zonal flow bands have similar
distributions in the spatial and temporal domains. 
\end{abstract}
\keywords{Solar activity, Helioseismology, Rotation, Magnetic fields}
\end{opening}

\section{Introduction}

 The pattern of temporal variations in the solar differential-rotation
rates discovered by Howard and LaBonte (1980)  from the Mt.~Wilson
full-disc Dopplergram data is known as torsional oscillations. These
manifest themselves as alternating latitudinal bands of slightly faster and slower
than the average rotation velocities migrating from pole to the Equator
in about 22 years. The low velocity of  3\,--\,5 m s$^{-1}$ of these zonal
flows makes it difficult to detect them from the global rotation signal
which is more than two orders of magnitude stronger, and it was remarkable
that Howard and LaBonte were able to isolate these weak signals. The close
correspondence between the torsional oscillation pattern and the surface
magnetic-flux distribution in the sunspot latitudes led Howard and LaBonte
(1980)  to conclude that this velocity field is perhaps, the signature
of a ``large-scale deep-seated phenomenon'' and that ``this
velocity field is associated in some way with the subsurface magnetic
fields that are responsible for the solar cycle''.

A subsequent
investigation by Snodgrass and Howard (1985) used the corrected values
for the coefficients $A$, $B$, and $C$ of the parabolic fit for the global
rotation to show that the full torsional oscillation pattern is not in the
form of a continuous wave running from poles to the Equator but rather
consisted of a high-latitude and a low-latitude branch with a break around
40\,--\,$50^\circ$ with each of the two components
consisting of a fast and a slow stream alternating with each other in
time.  Snodgrass and Howard (1985) also noticed that the low-latitude
stream of enhanced shear that lies in between the fast zone and its
poleward adjacent zone spatially coincides with the centroid of the
sunspot distribution.  Later, Snodgrass (1987, his Figure 2)  in
addition to confirming the overlap of the shear enhanced pattern with the
sunspot distribution (referred to as the Butterfly pattern)  for
cycles 20 and 21, demonstrated that the zone of diminished shear is
located in between the two successive sunspot activity zones, {\it i.e.}, in
between two successive butterfly patterns. This figure also showed
that there are no magnetic features corresponding to the shear increase
and decrease zones at latitudes beyond $\pm 60^\circ$. Makarov and
Sivaraman (1989) in fact, suggested that these zones of increased or
decreased shear at high latitudes correspond spatially with the
polar faculae distribution. Subsequent painstaking efforts by Ulrich
\etal~(1988), Snodgrass (1992), and Ulrich (2001) and the references therein)
refined the methods of reduction of the Mt.~Wilson full-disc velocity maps,
leading to a considerable improvement in the visibility of torsional-oscillation
signal against the background noise and, indeed, established
the reality of the existence of this velocity pattern on the solar surface.

      All of the above-mentioned works refer to the rotation rate near the
solar surface. With the advent of helioseismology it has become possible
to estimate the rotation rate in the solar interior by inversions of the
rotational splittings of the solar-oscillation frequencies from the
accurately measured helioseismic data obtained by the ground-based Global
Oscillation Network Group (GONG) and by the Michelson Doppler Imager (MDI)
onboard the SOHO spacecraft.  The rotation-rate residuals derived
by subtracting the time-averaged rotation rate from the corresponding
rotation rate at each depth and latitude show temporal variations with
faster and slower rotating bands moving equatorward with time (Schou 1999;
Howe \etal~2000; Antia and Basu 2000). We refer to this as the zonal-flow
pattern in this paper.  This pattern is very similar to the torsional
oscillation bands observed on the surface (Howard and LaBonte 1980;  
Ulrich \etal~1988; Snodgrass 1992; Ulrich 2001) but because of smaller errors in the
seismic data, it is more well defined and robust than the latter. Further
investigations by Antia and Basu (2001), Vorontsov \etal~(2002), Basu
and Antia (2003), Howe \etal~(2006) using more extensive data from GONG
and MDI have revealed results sufficient to build a fairly consistent
picture of the time-dependent structure and dynamics of these zonal flows
at different depths and in different latitudes in the solar interior. The
patterns derived from the GONG and MDI data are generally in reasonably
good agreement with each other.  Finally, the results from the recent
study by Basu and Antia (2006) and Howe \etal~(2006) using data
covering almost a complete sunspot cycle (1996 to 2006)  have
consolidated the properties of the zonal flows and provided a fairly
complete picture of these migrating zonal bands.

The zonal-flow pattern in
the solar interior has now become available for almost the full sunspot
cycle from helioseismic data. It would therefore be of interest to compare
the zonal-flow pattern with the distribution of surface magnetic features
which are manifestations of the cyclic magnetic activity and to look for
similarities and differences between them. With this aim, we plot the
patterns of distribution of surface magnetic features,
namely sunspots and the polar faculae, on zonal-flow contour maps in the
subsurface region at $r=0.98R_\odot$ derived from the GONG data for the
period 1995\,--\,2007.

  The rest of the paper is organised as follows: In Section 2 we describe
the data and the analysis procedure, in Section 3 we present the results,
and finally in Section 4 we summarise the conclusions.

\section{Data and  Analysis}

\subsection{Helioseismic Data and the Inferred Zonal Flow Pattern }

In order to infer the rotation rate in the solar interior we use 120
temporally overlapping data sets from GONG (Hill \etal~1996) each
covering a period of 108 days from 7 May 1995 to 15 May 2007, with a
spacing of 36 days between successive data sets. Each data set consists of
the mean frequencies of different $(n,\ell)$ multiplets and the splitting
coefficients. We use the 2D Regularised Least Squares (2DRLS) inversion
technique (Antia, Basu, and Chitre~1998)  for deducing the rotation rate for each of
these data sets. We take the temporal average over all of these data sets to
find the mean rotation rate at each latitude and depth covered in the
study. The temporal mean is subtracted from the rotation rate at each
epoch to find the residuals ($\delta\Omega$) which give the temporally varying
component of the rotation rate. This is referred to as the zonal flow.
Thus we have,
\begin{equation}
\delta\Omega(r,\theta,t)=\Omega(r,\theta,t)-\langle \Omega(r,\theta,t)\rangle,
\end{equation}
where $\Omega(r,\theta, t)$ is the rotation rate
as a function of the radial distance $r$, latitude $\theta$, and time $t$.  The
angular brackets denote the temporal average over the data ensemble. It
should be noted that the splitting coefficients are sensitive only to the
North\,--\,South symmetric component of rotation rate and hence the inferred
rotation pattern always looks symmetrical about the Equator. The actual
zonal flow pattern may have some asymmetry which cannot be detected from
the seismic data used in this study.

The zonal flow pattern consisting of fast-rotating streams
(represented by contours in red) and slow-rotating
streams (represented by contours in blue) is shown in Figure~1.  At low
latitudes this flow pattern consists of bands of fast- and slow-moving
fluid which move towards the Equator with time in both hemispheres, which
eventually meet near the Equator.  At high latitudes, the behaviour of zonal
flow pattern is quite different as the bands of fast- and slow-moving
streams migrate polewards with time. The slow stream reaches the poles
near the time of polar-field reversal around 2000.7. The zonal flow
velocities are rather small around latitudes of 40\,--\,$50^\circ$ (Vorontsov
\etal~2002; Basu and Antia 2003). This region acts as a boundary separating
the zonal-flow systems at low and high latitudes. Snodgrass and Howard
(1985) also found a break in the surface torsional oscillation pattern around
the same latitudes.

\begin{figure}
\centerline{\includegraphics[width=0.9\textwidth]{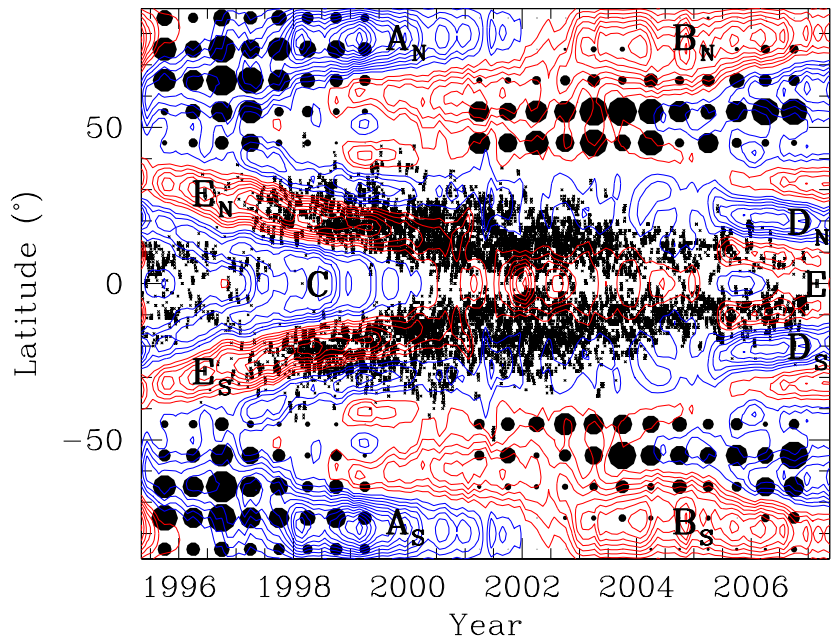}}
\caption{Contours of constant residual rotation velocity,
$\delta v_\phi=\delta\Omega r\cos\theta$, at $r=0.98R_\odot$
obtained from GONG data are shown as a function of time and latitude.
The red contours show positive values while the blue contours show negative
values. The contour spacing is 1~m\,s$^{-1}$ and the zero contour is not shown.
Different bands of faster ({\it e.g.}, $B_N$, $B_S$, $E$, $E_N$, $E_S$) and
slower ({\it e.g.}, $A_N$, $A_S$, $C$, $D_N$, $D_S$) than average rotation velocity are
marked in the figure.
The markings at low latitudes show the position of sunspots.
The filled black circles at
high latitudes represent the daily mean number of polar faculae
averaged over a period of six months in $10^\circ$ latitude bins
from $40^\circ$ to $90^\circ$.
The area of the circles is proportional to the number of PFs.}
\end{figure}
 
\subsection{Properties of  Polar Faculae}

\subsubsection{Number Counts and Latitudinal Distribution }

           Soon after the polar-field reversal, which occurs after
sunspot activity has reached its peak, polar faculae (abbreviated as PF)  
make their appearance in the latitude zones poleward of $40^\circ$
in both the hemispheres.  They can be identified easily on the
broad-band images as tiny bright features of sizes ranging from 3\,--\,$10''$
with a contrast $I_{\rm PF}/I_{\rm photosphere}  = 1.03$ to 1.10 and on
the K-line spectroheliograms as bright points located preferentially along
the network boundaries {\it i.e.}, along the boundaries of the supergranular
cells (Makarov and Makarova 1996).  The cyclically varying numbers of PF,
which are $180^\circ$ out of phase with the sunspot number, over the entire
solar disk are highly correlated with the polar magnetic-field strengths
(Sheeley 1964, 1976, 1991).  The total counts and the latitude
distribution of PF determined for the period 1940\,--\,1985 by Makarov,
Makarova, and Sivaraman (1989), Makarov and Sivaraman (1989), Makarov
and Makarova (1996) have been extended to the present day
by Makarov, Makarova, and Callebaut (in preparation).
The procedure that was adopted for
determining the total counts, as well as the latitude distribution of the
PF, is described in Makarov and Makarova (1996) and may be summarised as
follows:  For every month, about 15 high-quality broad-band images were
selected from the Kislovodsk Solar Station collections. The gaps in the
data were filled in using the Ca~{\sc ii}~K spectroheliograms of the Kodaikanal
Observatory collections.  The numbers of PF within the latitude zones
starting from $40^\circ$ and reaching up to the poles were counted in steps of
$10^\circ$ in order to derive their distribution. This was
done plate by plate for the northern and southern hemispheres separately
and from the counts so obtained for every month, the monthly and the
six-monthly averages were worked out.  We use the six-monthly mean values of
PF in each of the $10^\circ$ bins for comparison with the zonal-flow pattern
in Figure~1. These are represented by filled circles with area proportional to
the six-monthly means plotted at the mean latitude positions of each of the
$10^\circ$ bins.

The PFs appear first at latitude zones $\pm 40$\,--\,$60^\circ$. The zones of
appearance of the PF expand and reach almost to the poles filling the
entire high-latitude regions during the sunspot minimum years. On
average, about 900 PF emerge per day during this epoch. The zones of PF
are seen to shrink progressively in extent during the rising phase of the
then-current sunspot cycle until they finally disappear with the polar
field reversal around the maximum phase (Figure 4 of Makarov and
Sivaraman 1989). {Sheeley and Warren (2006) used images and
magnetograms from MDI
to study PFs during the period 1996\,--\,2005 and showed that in each hemisphere
there is a facula-free zone separating the old-cycle polar field
from trailing-polarity flux that is migrating poleward from the sunspot
belts. These facula-free zones coincide with the neutral lines of the
axisymmetric component of photospheric magnetic field and their arrival
at the poles in 2001 marks the reversal of the polar fields.}

\subsubsection{Magnetic Fields of Polar Faculae as the Source of Polar Fields}

   PFs have most commonly sizes in the range of 3\,--\,$10''$ when they
occur as individual structures or as bipoles, although a few of them
appear at smaller sizes reaching down to  $\approx1''$.  A few others
showing a complex structure have sizes exceeding $10''$ and at times
even as high as $30''$ (Makarov and Makarova 1996). In the
magnetograms they appear in the form of flux knots either bright or dark
depending on their magnetic polarity. Based on the polarimetric
measurements by Homann, Kneer, and Makarov~(1997), Makarov and Makarova (1998)  
estimate the magnetic flux per faculae to be $\approx 7\times10^{19}$ Mx, while Varsik, Wilson, and Li
(1999) estimate the flux to be $\approx 10^{19}$ Mx from calibrated
low-resolution magnetograms. Thus the PFs possess a range of magnetic-field
strengths from 150 to 1700 G depending on both their sizes and
the amount of flux they carry, although PF with low field strengths
appear to be more common.

 The evolution of the integrated flux over the polar regions was traced by
Lin, Varsik, and Zirin~(1994) using high resolution magnetograms covering the period
from early-1991 to mid-1993 that spans the maximum and declining phases of
the sunspot cycle 22. According to them, during the solar-maximum phase,
the polar regions are populated by magnetic elements of positive and
negative polarity of almost equal numbers and of equal field strengths
rendering the net fields at the poles close to zero.
This presumably represents the polar-field reversal epoch. With the
progress of the sunspot cycle towards the minimum, the elements of one
polarity outnumber those of the opposite polarity in terms of the field
strengths and numbers rendering the fields at the poles predominantly of
one polarity.  A subsequent study using more of such high-resolution
magnetogram sequences during sunspot minimum phase by Varsik, Wilson,
and Li~(1999)
confirms that knots of one polarity far exceed those of opposite
polarity. It is this excess of one polarity flux elements over the other,
itself varying cyclically (from + to $-$ and then to + and so on), that
decides the polarity of the field at the poles of the Sun in any given
cycle. It is the collective net field of the PF that a magnetograph
operating at low resolution measures as the magnetic flux at the poles of
the Sun during the sunspot minimum phase.

\section{Results}

\subsection{Latitude Distribution of Sunspots and the Zonal Flow Pattern}

We have shown in Figure~1 the latitudinal distribution of sunspots (the
Butterfly pattern) extracted from the Greenwich sunspot data
superposed on the plot of the zonal-velocity band pattern at $r=0.98R_\odot$.
We find that the dense part of the distribution of spots lies over the
zonal bands of the faster than average rotation rate ($E_N$ and $E_S$) while,
the less dense part of the distribution spills over to the slow streams
($D_N$ and $D_S$). There is also a sprinkling of sunspots in the region C
(the slow stream). These might possibly be very small spots or pores being
the last vestiges of solar cycle 22 (1986\,--\,1996).  Similar spatial
coincidence between sunspot distribution and surface torsional oscillation
pattern has been noted earlier by Snodgrass (1987).

\subsection{PF Distribution and the High Latitude Zonal Flow Pattern }

It is evident from Figure~1 that the polar-faculae distribution coincides
well, both spatially and temporally, with the high-latitude zonal-band
pattern ($A_N$ in the north and $A_S$ in the south hemispheres). Both the
slow-stream bands $A_N$ and $A_S$ and the PF associated with them that have
reached close to the respective poles disappear with the polar-field
reversal in 2000.7. Two new zonal bands of fast streams ($B_N$ and $B_S$)
that originated at $\pm 50^\circ$ latitudes about two years prior to the
polar-field reversal, have now ascended poleward filling the latitude
regions where $A_N$ and $A_S$ were present before the polar-field reversal.
The PF of the new cycle (2002\,--\,2006) that appeared soon after the
polar-field reversal coinciding with the new fast streams $B_N$ and $B_S$ have
also migrated polewards synchronously with the fast streams.  The polar-field
reversal in 2000.7 that marks the end of one PF cycle and the
beginning of the next one in both hemispheres is also the epoch that
marks the end of the slower-than-average rotation-rate zonal bands at the
poles ($A_N$ and $A_S$)  and the ascendancy of the new faster than average
rotation rate zonal bands ($B_N$ and $B_S$) to their positions. Thus the
distribution of polar faculae that provides the magnetic field at the
poles shows a distribution in latitude and time similar to the zonal flow
pattern of the sub-surface layers at $r=0.98R_\odot$ at high
latitudes. Although, we have used the zonal-flow pattern at $r=0.98 R_\odot$ for
establishing the spatial and temporal coincidence with the PF, similar
correspondence should hold for all sub-surface layers lying above
$r=0.95 R_\odot$. Below $r=0.90R_\odot$ the zonal-flow pattern
appears to be smeared out and the phase also changes at low
latitudes. Thus at $r = 0.8 R_\odot$ the fast and slow streams can
hardly be recognised at the high latitudes (Antia and Basu 2001, Figure~4;  
Basu and Antia 2006, Figure~1). At any rate it appears that the correspondence
between zonal flow pattern and the PF seems to be confined to the layers
above $r = 0.95 R_\odot$.

\section{Discussion and Conclusions}

We have derived the residual rotation rates by subtracting the time-averaged
rotation rate from that at each epoch from helioseismic data from the
GONG project for a full solar cycle (1996\,--\,2007) at $r = 0.98 R_\odot$ in the
solar interior. These show a set of zonal velocity bands moving with
faster- and slower-than-average rotation rate. The zonal velocity bands
have two components per hemisphere (Figure~1) (a) the high-latitude
component of alternating slow (blue contours) and fast (red contours)  
streams (above $\approx 50^\circ$) that move poleward ($A_N$ or $A_S$ and $B_N$ or $B_S$),
(b) the low-latitude component of fast ($E_N$ or $E_S$) and slow ($D_N$ or $D_S$)
streams in the sunspot latitudes that move towards the Equator.  $E_N$
and $E_S$ later merge to form a single fast stream $E$.  From the earlier
studies (Vorontsov \etal~2002; Basu and Antia 2003) it is known that the
zonal flows in the high-latitude as well as in the sunspot-latitude belts
persist throughout much of the convection zone and are quite stable.  The
$\approx 50^\circ$ latitude in the two hemispheres seems to be the boundary that
separates the high latitude streams from the low latitude streams.
Interestingly, this is also the latitude region where the rotation rate
residual is close to zero throughout almost the entire convection zone
(Vorontsov \etal~2002; Basu and Antia 2006).

We have established that (a)  polar faculae and the zonal-flow bands in the
region above  $\pm 40$\,--\,$50^\circ$ latitudes have very similar distribution
in the spatial and temporal domains, irrespective of whether the zonal
velocity band is a fast stream or a slow stream (Figure~1). The switch from
fast to slow happens around sunspot minimum (around 1996) when
there is no reversal in polarity of the polar magnetic field, while the
switch from slow to fast occurs around the sunspot maximum which coincides
with field polarity reversal at the poles around 2000.7. Thus
there is one pair of streams (one fast and one slow)  during the period
of two successive polar-field reversals and both components of the pair
are associated with polar fields of the same polarity either positive or
negative as the case might be. (b)  In the sunspot latitudes, the
butterfly pattern coincides with the fast streams ($E_N$, $E_S$ or $E$)
although part of the sunspot distribution spills over to the slow streams
($D_N$ and $D_S$) too (Figure~1).
{Of course, our study is restricted
to one solar cycle for which the seismic data is available and this
association needs to be confirmed in subsequent cycles.}
          
We have also established that there is a striking similarity in spatial
and temporal organisation between the zonal-flow streams in the interior
and the surface magnetic fields.
This would imply a close
coupling between the periodic components of flows in the interior
and the surface magnetic-field structures which are visible manifestations
of the cyclic magnetic activity. This close similarity raises interesting
possibilities about their connection links ({\it cf.}, Snodgrass 1987).  
Clearly, the velocity changes present in the zonal flow alone are
intrinsically too weak to drive the global solar-activity cycle. It is,
therefore, possible that both the zonal-flow pattern and the overall solar
magnetic activity are manifestations of a common coherently-driven global
mechanism that remains to be identified and properly understood.  There
could be other possibilities:  two mechanisms operating on disparate
scales at two different depths in the convection zone could conceivably
result in mutually-coherent velocity and magnetic field patterns.  It is
commonly accepted that the shear zone below the base of the convection
zone is the seat of the dynamo that amplifies and produces the strong
magnetic field that gives rise to sunspots.  Likewise, there is an
amplification of the magnetic field by the shear in the sub-surface layers
between $r=0.98R_\odot$ and $0.95R_\odot$ that could produce weaker fields
like those in the polar faculae (Dikpati \etal~2002).
The possible role of the near-surface shear layer in small
scale amplification of magnetic field has been envisaged earlier by Gilman
(2000). More sophisticated theoretical models supported by simulations
to explore the mechanisms that can generate and sustain the zonal-band
systems in the interior and also organise the magnetic fields in a
mutually coherent way are clearly needed. It is gratifying to note that
initial efforts in this direction have already produced migrating patterns
({\it e.g.}, Covas \etal~2000; Covas, Tavakol, and Moss 2001; Covas,
Moss, and Tavakol 2004; Lanza 2007).  It would equally be
important to explore the role of the subsurface shear region in
small-scale amplification of magnetic fields and to explain the intimate
relationship between the family of zonal-band system and the distribution
of magnetic flux elements.

The helioseismic data from GONG and MDI
accumulated over the past solar cycle 23 enabled Antia, Chitre, and Gough
(2008) to study temporal variations in the solar rotational kinetic-energy.
It was demonstrated that at high latitudes ($> 45^\circ$) 
variation in the kinetic energy through the convection zone is correlated
with the solar activity, while in the equatorial latitudes ($< 45^\circ$)
it is anticorrelated except for the upper 10\% of the solar radius where
both are in phase. The amplitude of temporal variation of the rotational
kinetic-energy integrated over the entire convection zone turns out to be
$\approx 3\times 10^{38}$ ergs implying a rate of variation of about
$5\times10^{30}$ ergs s$^{-1}$
over the solar cycle. From energy conservation it is
expected that the torsional kinetic-energy variation is comparable with
that in the magnetic energy but with opposite phase.  It thus seems that
the temporal variation in rotational kinetic energy in the convection zone
is related to the solar cycle with its tantalising similarity with the
magnetic activity cycle.

\begin{acks}
This work  utilises data obtained by the Global Oscillation
Network Group (GONG) project, managed by the National Solar Observatory,
which is
operated by AURA, Inc. under a cooperative agreement with the
National Science Foundation. The data were acquired by instruments
operated by the Big Bear Solar Observatory, High Altitude Observatory,
Learmonth Solar Observatory, Udaipur Solar Observatory, Instituto de
Astrofisico de Canarias, and Cerro Tololo Inter-American Observatory.
We thank Baba Varghese for his valuable help in formulating the Figure.
S.M.C. thanks the Indian National Science Academy for support under the
INSA Honorary Scientist programme.
\end{acks}

\end{article}
\end{document}